\documentclass[prb,aps,amssym,nofootinbib,floatfix,reprint,notitlepage]{revtex4-2} 

\usepackage[pdfpagelabels,colorlinks=true,breaklinks=true,linktocpage=true,linkcolor=red,urlcolor=blue,citecolor=blue,%
]{hyperref}

\usepackage{tabularx}
\usepackage{tabto}
\usepackage{mathtools}
\usepackage{tensor}
\usepackage{xcolor}
\usepackage{amsmath,amssymb,bbold,bm}
\usepackage{graphicx}
\usepackage[normalem]{ulem}
\usepackage{mathdots}
\usepackage{caption}
\usepackage{subcaption}
\usepackage{cancel}
\usepackage{bm}
\usepackage{comment}
\usepackage{currfile}
\usepackage{tikz}
\usepackage[compat=1.1.0]{tikz-feynman}
\usepackage{feynmp}
\usepackage{tikz-feynhand}
\usetikzlibrary{arrows.meta, positioning}

\usepackage{esint}
\usepackage{empheq}

\usepackage[most]{tcolorbox} 
\newtcolorbox{proofbox}{
    colback=gray!15, 
    colframe=gray!50, 
    boxrule=0.5pt,    
    arc=3mm,          
    left=5pt,         
    right=5pt,        
    top=5pt,          
    bottom=5pt        
}

\usepackage[justification=raggedright,singlelinecheck=false]{caption} 

\DeclareDocumentCommand\vb{ s m }{\IfBooleanTF{#1}{\boldsymbol{#2}}{\mathbf{#2}}} 

\begin{document}
\newcommand{\be}{\begin{equation}}
\newcommand{\ben}{\begin{equation*}}
\newcommand{\ee}{\end{equation}}
\newcommand{\een}{\end{equation*}}
\newcommand{\bs}{\begin{split}}
\newcommand{\es}{\end{split}}
\newcommand{\bmx}{\begin{array}}
\newcommand{\emx}{\end{array}}
\newcommand{\bea}{\begin{eqnarray}}
\newcommand{\bean}{\begin{eqnarray*}}
\newcommand{\eea}{\end{eqnarray}}
\newcommand{\eean}{\end{eqnarray*}}
\newcommand{\dg}{^{\dagger}}
\newcommand{\dn}{^{\vphantom{\dagger}}}
\newcommand{\nn}{\vphantom{-}}
\newcommand{\kv}{{\bf k}}
\newcommand{\rv}{{\bf r}}
\newcommand{\lr}{\leftrightarrow}
\newcommand{\ra}{\rightarrow}
\newcommand{\la}{\leftarrow}
\newcommand{\ua}{\uparrow}
\newcommand{\da}{\downarrow}
\newcommand{\Ua}{\Uparrow}
\newcommand{\Da}{\Downarrow}
\newcommand{\bb}[1]{\mathbb{#1}}
\newcommand{\qqquad}{\qquad\qquad\qquad}
\newcommand{\so}{\qquad\rightarrow\qquad}
\newcommand{\So}{\qquad\Rightarrow\qquad}
\newcommand{\smallnote}[1] {\begin{flushleft} \rule{10cm}{0.4pt} \end{flushleft}{\small #1}\begin{flushright} \rule{10cm}{0.4pt} \end{flushright}}
\newcommand{\orr}{\qquad\text{or}\qquad}
\newcommand{\andd}{\qquad\text{and}\qquad}
\newcommand{\lvec}{\overleftarrow}
\newcommand{\eps}{\epsilon}
\newcommand{\tq}{\tilde{q}}
\newcommand{\tpsi}{\tilde{\psi}}
\newcommand{\veps}{\varepsilon}
\newcommand{\tphi}{\tilde{\phi}}
\newcommand{\ocite}[1]{[\onlinecite{#1}]}
\newcommand{\sgn}[1]{{\rm sign}{#1}}
\newcommand{\pref}[1]{(\ref{#1})}
\newcommand{\tildr}[2]{\intopi{d{#1}}\Big(g_{#1}^R(\eps_p){#2}\Big)}
\newcommand{\tilda}[2]{\intopi{d{#1}}\Big(g_{#1}^A(\eps_p){#2}\Big)}
\newcommand{\dod}[1]{\frac{d}{d#1}}
\newcommand{\dodd}[1]{\frac{d^2}{d#1^2}}
\newcommand{\intinf}[1]{\int_{-\infty}^{+\infty}{#1}}
\newcommand{\intoinf}[1]{\int_{0}^{\infty}{#1}}
\newcommand{\intopi}[1]{\int_{0}^{\pi}{#1}}
\newcommand{\intbb}[1]{\int_{-\beta}^{\beta}{#1}}
\newcommand{\intob}[1]{\int_{0}^{\beta}{#1}}
\newcommand{\intpi}[1]{\int_{-\pi}^{+\pi}{#1}}
\newcommand{\suminf}[1]{\sum\limits_{#1=-\infty}^{+\infty}}
\newcommand{\sumoinf}[1]{\sum\limits_{#1=0}^{+\infty}}
\newcommand{\sumvinf}[1]{\sum\limits_{#1=1}^{+\infty}}
\newcommand{\re}[1]{{\rm Re}\left[ #1 \right]}
\newcommand{\im}[1]{{\rm Im}\left[ #1 \right]}
\newcommand{\tr}[1]{{\rm Tr}\left[ #1 \right]}
\newcommand{\ev}[1]{{\rm Even}\left[#1\right]}
\newcommand{\od}[1]{{\rm Odd}\left[#1\right]}
\newcommand{\pov}[1]{{\rm P}\frac{1}{#1}}
\newcommand{\fo}[1]{\frac{1}{#1}}

\newcommand{\abs}[1]{\left\vert #1 \right\vert}
\newcommand{\bra}[1]{\left\langle #1 \right\vert}
\newcommand{\ket}[1]{\left\vert #1\right\rangle}
\newcommand{\braa}[1]{\left\langle\left\langle #1 \right\vert\right.}
\newcommand{\kett}[1]{\left.\left\vert #1\right\rangle\right\rangle}
\newcommand{\braket}[1]{\left\langle #1\right\rangle}
\newcommand{\brket}[1]{\langle #1\rangle}
\newcommand{\dbraket}[1]{\left\langle\langle #1\right\rangle\rangle}
\newcommand{\com}[2]{\left[#1,#2\right]}
\newcommand{\acom}[2]{\left\{#1,#2\right\}}
\newcommand{\mat}[1]{\left(\bmx{cc}#1\emx\right)}
\newcommand{\matc}[2]{\left(\bmx{#1}#2\emx\right)}
\newcommand{\matlc}[2]{\bmx{#1}#2\emx}
\newcommand{\matn}[1]{\bmx{cc}#1\emx}
\newcommand{\matl}[1]{\bmx{ll}#1\emx}
\newcommand{\diag}[2]{\left(\bmx{cc}#1 & 0 \\ 0 & #2\emx\right)}
\newcommand{\paulix}{\mat{0&1\\1&0}}
\newcommand{\pauliy}{\mat{0&-i\\i&0}}
\newcommand{\ipauliy}{\mat{0&1\\-1&0}}
\newcommand{\pauliz}{\mat{1&0\\0&-1}}
\newcommand{\mpauliz}{\mat{-11&0\\0&1}}
\newcommand{\sepline}{\begin{center}\rule{8cm}{.5pt}\end{center}}
\newcommand{\hi}{\noindent\currfilebase.pdf \hspace{.2cm}-\hspace{.2cm} \today}
\newcommand{\nq}{\mathsf}
\newcommand{\bnq}[1]{\breve{\mathsf #1}}
\newcommand{\bv}[1]{\breve{#1}}

\setlength{\parindent}{0.5cm}
\newcommand{\indentoff}{\setlength{\parindent}{0cm}}

\newcommand{\red}[1]{{\color{red} #1}}
\newcommand{\yellow}[1]{{\color{yellow} #1}}
\newcommand{\green}[1]{{\color{green} #1}}
\newcommand{\blue}[1]{{\color{blue} #1}}
\newcommand{\black}[1]{{\color{black} #1}}
\newcommand{\purple}[1]{{\color{purple} #1}}
\newcommand{\brown}[1]{{\color{brown} #1}}
\newcommand{\ygg}[1]{{\color{orange} #1}}
\newcommand{\yg}[1]{{\color{teal} #1}}
\newcommand{\nothing}[1]{}

\newcommand{\sbraket}[1]{\langle #1 \rangle}
\newcommand{\brk}{\newline\noindent\rule{\textwidth}{0.4pt}\newline}
\newcommand{\vk}{\vb{k}}

\makeatletter
\newcommand{\setword}[2]{%
  \phantomsection
  #1\def\@currentlabel{\unexpanded{#1}}\label{#2}%
}
\makeatother

\title{ Vestigial $d$-wave charge-$4e$ Superconductivity from Bidirectional Pair Density Waves}
\author{Ethan Huecker}
\author{Yuxuan Wang}
 \affiliation{Department of Physics, University of Florida, Gainesville, Florida, 32607, USA}
\date{\today}
\begin{abstract}
	We analyze the leading vestigial instability due to the melting of a bidirectional pair-density-wave state in two dimensions. In a previous work by one of the authors \cite{YuxuanandWu2024}, it was found that the interplay between pair-density-wave fluctuations with ordering momenta along the $x$ and $y$ directions can provide a strong attractive interaction for charge-$4e$ superconductivity in the $d$-wave channel. In this work, we go beyond the artificial large-$M$ mean-field theory previously adopted and compute the phase diagram by incorporating phase fluctuations of the pair-density-wave order parameters. By investigating the relevance of various topological defects, we show that the interaction in the $d$-wave channel, together with the strong anisotropy of phase fluctuations around the pair-density-wave ordering momenta, favors a vestigial charge-$4e$ superconducting order at intermediate temperatures. By contrast, a competing charge-density-wave vestigial order does not develop, due to the suppression of its stiffness.
\end{abstract}
\maketitle

\section{Introduction}

Charge-$4e$ superconductivity ($4e$-SC) is an unconventional superconducting phase in which the charge carriers forming the superfluid consist of quartets of electrons, as opposed to Cooper pairs in usual superconductors ($2e$-SC). In addition to sharing characteristic features of $2e$-SC~\cite{Gnezdilov2022Charge4e}, such as the Meissner effect and zero resistance, $4e$-SC supports fractional topological defects, i.e., half vortices with a half flux quantum $\Phi_0/2=hc/4e$. Such signatures have been observed in Little-Parks oscillations in the kagome metal $\mathrm{CsV}_3\mathrm{Sb}_5$~\cite{Ge2022Charge4e6e} (along with more prominent signatures for $6e$-SC) and in superconducting quantum interference devices (SQUIDs)~\cite{Ciaccia2024Charge4e} fabricated from a two-dimensional InAs-Al superconductor-semiconductor heterostructure.


Unlike $2e$-SC, $4e$-SC does not develop via a weak-coupling instability of a Fermi surface; identifying its microscopic mechanism in strongly correlated systems is a nontrivial problem \cite{Samoilenka2025MicroscopicElectronQuadrupling,Chirolli2024CooperQuartets,Soldini2024Charge4eHubbard}.
A promising avenue has been found in systems with intertwined orders, where multiple symmetries broken by some primary order at low temperature are restored sequentially by thermal fluctuations \cite{Fradkin2015,fradkin2024,Babaev2004_defect,Herland2010MetallicSuperfluidCharge4e,Bojesen2013TimeReversalSymmetryBreakdown}.
In these systems, $4e$-SC can emerge as a vestigial order from fluctuations of multi-component pairing order parameters, i.e., $\Delta_{4e}\sim \Delta_1 \Delta_2$. Individually, $\Delta_{1,2}$ break point-group symmetries~\cite{Fernandes2021Charge4e,Jian2021Charge4e, hecker, Hecker2023CascadeVestigial,scheurer-1,scheurer-2,Alkady2025SymmetryInducedPairing} or translation symmetries~\cite{Berg2009,Agterberg2011Charge6e,Radzihovsky2009QLC,Volovik2024FermionicQuartet,Agterberg2008}, in addition to the usual $U(1)$ symmetry.
Within mean-field theory, once the pairing orders develop, all symmetries are broken. However, when fluctuation effects are included, the system may exhibit broken $U(1)$ symmetry\footnote{In two spatial dimensions, by ``broken symmetry'' we mean that the system is in a phase of quasi-long-range order, with algebraically decaying correlations.} while crystalline symmetries remain intact: in such a state $\Delta_{4e}$ becomes (quasi-)long-range ordered while $\Delta_{1,2}$ remain disordered. Indeed, such a scenario has been investigated in Fulde-Ferrell-Larkin-Ovchinnikov (FFLO) superconductors~\cite{Radzihovsky2009QLC,Agterberg2011Charge6e,radzPRA}, multi-component pairing on a hexagonal lattice~\cite{Fernandes2021Charge4e,Hecker2023CascadeVestigial}, and in pair density wave (PDW) states~\cite{Berg2009,Agterberg2008}. However, one key challenge in this route is that $4e$-SC is generally in competition with other vestigial orders, all of which need to be treated on equal footing. For example, in the case of multi-component pairing on a hexagonal lattice~\cite{Fernandes2021Charge4e,Hecker2023CascadeVestigial}, it was found that $4e$-SC is always secondary to a vestigial nematic order.

In this work, we focus on PDW order as the primary order, which is a $2e$ strongly-correlated superconducting state that breaks translation symmetry in the absence of a magnetic field or fine-tuned parameters~\cite{annurev:/content/journals/10.1146/annurev-conmatphys-031119-050711,WangChubukovAgterbergPDW,WangChubukovAgterbergPDW2,PhysRevLett.130.026001,PhysRevLett.131.026601,Setty2023,PhysRevB.107.214504,PhysRevB.110.094515,PhysRevLett.133.176501,PhysRevLett.125.167001,Huang2022,PhysRevB.109.L121101,wang2024pairdensitywavesstrongcoupling,PhysRevX.9.021047,PhysRevB.97.174511,wang2024quantumgeometryfacilitatedpairdensitywave,PhysRevB.105.L100509,PhysRevB.108.035135,Agterberg2008,Wu2024}. Recent experiments have reported possible signatures of PDW order in a variety of correlated electronic systems, including La-based underdoped high-$T_c$ cuprates~\cite{annurev:/content/journals/10.1146/annurev-conmatphys-031119-050711}, kagome metals~\cite{Chen2021,Deng2024,PhysRevB.108.L081117,PhysRevB.110.024501,yao2024selfconsistenttheory2times2pair,PhysRevLett.129.167001,Scammell2023,Song2025Phase,Lin2024Theory,ZhouWang2022_ChernFermiPocket}, NbSe$_2$~\cite{XiaolongLiu,Cao2024,PhysRevB.107.224516}, UTe$_2$~\cite{Gu2023,Aishwarya2023,Aishwarya2024}, EuRbFe$_4$As$_4$~\cite{Zhao2023}, SrTa$_2$S$_5$~\cite{Devarakonda2024}, and rhombohedral graphene~\cite{han2024signatureschiralsuperconductivityrhombohedral}.
In PDW states, $4e$-SC emerges as a vestigial order $\Delta_{4e} \sim \Delta_{\bm Q} \Delta_{-\bm Q}$, where $\Delta_{\pm\bm Q}$ are the PDW order parameters with Cooper-pair momenta $\pm \bm Q$, upon restoring translational symmetry. Indeed, $4e$-SC has been found to exist in a phenomenological model~\cite{Berg2009} for parameters within a certain range. Still, while the abundance of candidate materials for PDW suggests that $4e$-SC may exist in their phase diagrams, from the theoretical side the strong-coupling nature has so far prevented a controlled, microscopic analysis of vestigial orders. 

In bidirectional PDW systems with $C_4$ rotation symmetry~\cite{Agterberg2008}, a recent theoretical analysis by one of the authors~\cite{YuxuanandWu2024} has uncovered a strong attractive interaction toward $4e$-SC. In this system, the PDW order parameters carry momenta $\bm Q= (\pm Q,0)$ and $\bm{P}=(0,\pm Q)$, which we denote as $\Delta_{\pm \bm Q}$ and $\Delta_{\pm \bm P}$. Concretely, it was found that when a simple geometric relationship between the PDW wave vector and the Fermi momentum is satisfied, i.e., $|\bm Q| = \sqrt{2} k_F$, there exists a strong interaction among the PDW bosons that is most attractive in the ``Cooper'' channel with $d$-wave pairing symmetry. Through a mean-field theory, it was shown that upon lowering the temperature the system enters a $d$-wave $4e$-SC state $\Delta_{4e}\sim \Delta_{\bm Q}\Delta_{-\bm Q} - \Delta_{\bm P}\Delta_{-\bm P}$ via a thermal phase transition. Nevertheless, this approach is explicitly biased toward $4e$-SC---in fact the same bosonic interaction is also attractive in the charge-density-wave (CDW) channel~\cite{YuxuanandWu2024}, e.g., $\Psi_{\bm Q - \bm P}\sim \Delta_{\bm Q} \Delta_{\bm P}^*$, but such an instability was not treated on equal footing. Furthermore, the mean-field theory is blind to the true (infinite-order) Kosterlitz-Thouless (KT) transition characterizing the quasi-long-range order of the 2D model~\cite{Kosterlitz1973,Nelson1979,altland_simons_cmft}. A formal determination of the transition temperatures taking into account all fluctuations toward possible vestigial orders requires an approach beyond mean-field theory.

In this paper, we take the strong interaction in the $d$-wave $4e$-SC channel identified in Ref.~\cite{YuxuanandWu2024} as input, and extend the prior analysis through a non-linear sigma model (nLSM) of the phase fluctuations of the PDW order parameters. With the phase variables of the PDW order parameters as fundamental degrees of freedom, our analysis directly connects with the proliferation of topological defects that characterize the KT transitions. By mapping our nLSM to its associated sine-Gordon model, topological defects are represented by local vertex operators whose scaling dimensions govern the transition temperatures into the vestigial orders. This approach has been taken in previous works~\cite{Babaev2004_defect,Herland2010MetallicSuperfluidCharge4e,Bojesen2013TimeReversalSymmetryBreakdown,Jian2021Charge4e,Berg2009,Radzihovsky2009QLC,Agterberg2008,Agterberg2011Charge6e,radzPRA}, in our model the bidirectional nature of the PDW order and the strong interaction between $\Delta_{\pm\bm Q}$ and $\Delta_{\pm\bm P}$ greatly modify their dynamics. Via an analytic calculation, we find that when the anisotropy of PDW fluctuations around their ordering momenta is large enough, upon lowering the temperature there exists an intermediate $d$-wave $4e$-SC order in which $U(1)$ symmetry is broken while translation symmetry remains intact, shown in Fig.~\ref{fig:phasetrans}.

Although our results are inspired by a specific PDW model, we argue that the key ingredients of this mechanism of stabilizing $4e$-SC are quite general: the enhancement of the $4e$-SC stiffness from multiple patches of antipodal bosonic momenta, which is kinematically similar to the Cooper instability of a Fermi surface. Such enhancement is absent for the stiffness of CDW fluctuations, which is small (and highly anisotropic) by comparison. This small CDW stiffness is reminiscent of that exploited in \cite{Radzihovsky2009QLC,radzPRA} for the FFLO state, where the authors develop a similar nLSM. In fact, due to the microscopic $SO(2)$ rotation symmetry, the CDW stiffness is zero in the direction perpendicular to the ordering momentum, leading to the stabilization of a $4e$-SC vestigial order. As a key difference with our present work, the authors only considered FFLO order along one direction, which amounted to assuming a nematic order developing at much higher temperatures, and the enhancement of $4e$-SC from PDW fluctuations with different momenta was absent. It may be worthwhile to revisit that model and examine the existence of nematicity in FFLO states more carefully, as our results point to a further enhancement of $4e$-SC without the nematic order.

The rest of the paper is organized as follows. In Sec.~\ref{sec:model}, we develop a nLSM for the thermal fluctuations of the bidirectional PDW state, and perform a mapping to the dual sine-Gordon model. In Sec.~\ref{sec:analysis}, we systematically analyze this dual model and develop a phase diagram for the vestigial orders. We end in Sec.~\ref{sec:conclusion} with a few concluding remarks and an outlook. In the Appendices, we provide some details on the the attractive interaction toward $4e$-SC and the treatment of additional cross-gradient terms in the nLSM.  

\section{Model and Methods}
\label{sec:model}

In this section, we construct a long-wavelength nLSM for the thermal fluctuations of a bidirectional PDW state based on insight from the Ginzburg-Landau (GL) free energy obtained in Ref.~\cite{YuxuanandWu2024}. After shedding light on the various topological defects hosted by the bidirectional model, we relate the nLSM to an associated lattice XY model to facilitate a mapping to a dual sine-Gordon model that represents topological defects by local fields.

\subsection{Ginzburg-Landau Free Energy}

A PDW state with $C_4$ rotation symmetry is defined through the superconducting order parameter
\begin{equation}
	\Delta(\bm{x})=\sum_i\Delta_{\bm{Q}_i}(\bm{x})e^{i\bm{Q}_i\cdot\bm{x}},\label{OP}
\end{equation}
where $\bm{Q}_i\in\{\bm{Q},\bm{P},-\bm{Q},-\bm{P}\}$ denotes a set of PDW momenta related by $C_4$ rotation.
\begin{figure}

	\begin{center}
		\resizebox{0.45\textwidth}{!}{%
			\begin{tikzpicture}[
				every edge/.style={draw, -{Stealth[scale=1.5]}}
				]
                \def\PI{3.141592653589793}
                \def\TWOPITHIRD{2.094395102393195} 
                \def\scale{0.6}

                \fill[orange!60,opacity=0.7,rotate=45]
                plot[domain=-\TWOPITHIRD:\TWOPITHIRD,samples=400]
      ({\scale*\x},{\scale*(acos(0.5 - cos(\x r)) * pi/180)})
    -- 
    plot[domain=\TWOPITHIRD:-\TWOPITHIRD,samples=400]
      ({\scale*\x},{\scale*(-acos(0.5 - cos(\x r)) * pi/180)})
    -- cycle; 
				\draw[color=red!100, fill=red!30] (2,0) ellipse [x radius=0.075in, y radius=0.15in];
				\draw[color=red!100, fill=red!30] (0,2) ellipse [y radius=0.075in, x radius=0.15in];
				\draw[color=red!100, fill=red!30] (-2,0) ellipse [x radius=0.075in, y radius=0.15in];
				\draw[color=red!100, fill=red!30] (0,-2) ellipse [y radius=0.075in, x radius=0.15in];
				\node[fill=black, circle, inner sep=1pt] (K) at (2,0) {};
				\node[fill=black, circle, inner sep=1pt] (L) at (0,2) {};
				\node[fill=black, circle, inner sep=1pt] (M) at (-2,0) {};
				\node[fill=black, circle, inner sep=1pt] (N) at (0,-2) {};
				\path (M) edge[<->] (K);
				\path (N) edge[<->] (L);
				\node (E) at (1.5,0.25) {$\bm{Q}$};
				\node (F) at (0.25,1.5) {$\bm{P}$};
				\node (G) at (-1.55,-0.275) {$-\bm{Q}$};
				\node (H) at (-0.35,-1.5) {$-\bm{P}$};
				\node (I) at (1.1,1.1) {$\scriptstyle \bm{k}$};
				\node (J) at (-1.2,1.1) {$\scriptstyle -\bm{k}+\bm{P}$};
				\node (K) at (-1.5,-1.1) {$\scriptstyle \bm{k}-\bm{P}-\bm{Q}$};
				\node (L) at (1.15,-1.1) {$\scriptstyle -\bm{k}+\bm{Q}$};
				\node[fill=black, circle, inner sep=1pt] (M) at (0.885,0.885) {};
				\node[fill=black, circle, inner sep=1pt] (N) at (0.885,-0.885) {};
				\node[fill=black, circle, inner sep=1pt] (O) at (-0.885,0.885) {};
				\node[fill=black, circle, inner sep=1pt] (P) at (-0.885,-0.885) {};
				\node (Y) at (-3,0) {\scalebox{0.7}{$\sim$}$\,\frac{1}{\sqrt{\alpha_2}}$};
				\draw[decorate,decoration={brace,amplitude=6pt}] (-2.3,-0.35) -- (-2.3,0.35);
				\draw[decorate,decoration={brace,amplitude=6pt}] (1.8,0.5) -- (2.2,0.5);
				\node (Z) at (2,1.1) {\scalebox{0.7}{$\sim$}$\,\frac{1}{\sqrt{\alpha_1}}$};
                \node (Y2) at (3,0) {};
			\end{tikzpicture}
		}
	\end{center}
	\caption{\footnotesize Filled Fermi sea (orange) with $C_4$ rotation symmetry, and PDW momenta $\bm{Q}_i$ with internal fermions situated on the Fermi surface. Ellipses (red) represent anisotropic PDW fluctuations with longitudinal and transverse amplitudes that scale as $1/\sqrt{\alpha_1}$ and $1/\sqrt{\alpha_2}$ respectively.}
	\label{fig:FS}

\end{figure}
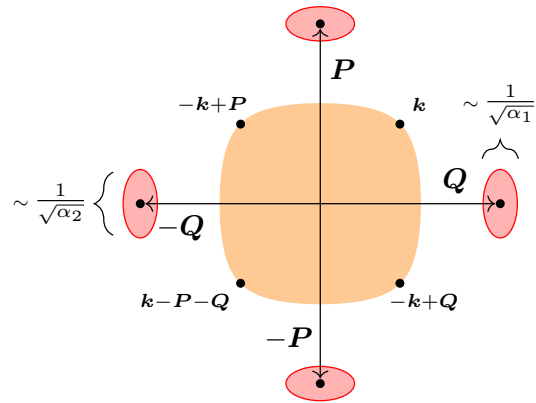
The GL free energy density for bidirectional PDWs was first studied in Ref.~\cite{Agterberg2008}, and further analyzed from the microscopic perspective in \cite{YuxuanandWu2024}, and takes the form
\begin{equation}
	\begin{split}
		\mathcal{F} & =\alpha\sum_i|\Delta_{\bm{Q}_i}|^2+\alpha_1\sum_i|\hat{\bm{Q}}_{i}^\parallel\cdot\boldsymbol{\partial}\Delta_{\bm{Q}_i}|^2 \\&\hspace{0.2in}+\alpha_2\sum_i|\hat{\bm{Q}}_{i}^\perp\cdot\boldsymbol{\partial}\Delta_{\bm{Q}_i}|^2\\&\hspace{0.2in}+\beta_1\sum_i|\Delta_{\bm{Q}_i}|^4+\beta_2|\Delta_{\pm\bm{Q}}|^2|\Delta_{\pm\bm{P}}|^2\\&\hspace{0.2in}+\beta\left(\Delta_{\bm{P}}\Delta_{-\bm{P}}\Delta^*_{\bm{Q}}\Delta^*_{-\bm{Q}}+\text{h.c.}\right)+\cdots,
	\end{split}\label{fmin}
\end{equation}
which respects the various parity, time-reversal, $U(1)$, and other point-group symmetries. We have incorporated the gradient terms $\alpha_{1,2}$, where $\hat{\bm{Q}}^\parallel_i$ and $\hat{\bm{Q}}^\perp_i$ are unit vectors parallel and perpendicular to the PDW momentum $\bm{Q}_i$. In general $\alpha_1\neq \alpha_2$. If the system has an $SO(2)$ rotation symmetry, then formally $\alpha_2 = 0$, as it would cost zero energy to vary the PDW wave vector in the direction $\hat{\bm{Q}}^\perp_i$~\cite{radzPRA,Radzihovsky2009QLC}; if the continuous rotation symmetry is only approximate, then $\alpha_1\gg \alpha_2$ (see Fig.~\ref{fig:FS}). The $\beta_1$ term is a simple repulsive self-interaction, and the $\beta_2$ term, when positive, captures a nematic instability.  The $\beta$ interaction can be decomposed via a Hubbard--Stratonovich transformation in the ``Cooper'' channel with $\sim \Delta_{\bm Q}\Delta_{-\bm Q}$. The positivity of $\beta$, which we show below, ensures that the latter interaction is attractive in the $d$-wave channel of the $4e$-SC sector described by the vestigial order parameter
\begin{equation}
	\Delta_{4e}=\Delta_{\bm{Q}}\Delta_{-\bm{Q}}-\Delta_{\bm{P}}\Delta_{-\bm{P}}.\label{rho4e}
\end{equation}
We note that within mean-field theory, the $\beta$ term can also be decomposed in the CDW channel as, e.g., $\Psi_{\bm Q - \bm P}\sim \Delta_{\bm Q} \Delta_{\bm P}^*$, and thus an unbiased treatment beyond mean-field theory is desired.

Interactions between PDW bosons are obtained by integrating out fermions, and as in Ref.~\cite{YuxuanandWu2024} we consider a special case in which $|\bm{Q}|/k_F=\sqrt{2}$, where $k_F$ is the Fermi momentum, shown in Fig.~\ref{fig:FS}.\footnote{For example, this condition is satisfied for PDW order in the spin-fermion model~\cite{WangChubukovAgterbergPDW1,WangChubukovAgterbergPDW2}. In the microscopic theory for PDW in Ref.~\cite{PDWorderelecrepul2023}, the ratio $|\bm{Q}|/k_F$ can be tuned.}  In this case, all fermions mediating the bosonic interactions can be placed on the Fermi surface at four hot-spots. By linearizing the fermionic dispersions around these hot-spots, a direct evaluation (see Appendix \ref{app:GL}) of the quartic-order coefficients yields
\begin{align}
	\beta_1\sim \frac{1}{v_F^2E_F}, &  & \beta_2\sim \frac{1}{v_F^2E_F}\ln\left(\frac{E_F}{T}\right), &  & \beta\sim \frac{1}{v_F^2T}.
\end{align}
It is clear that $\beta\gg\beta_{1},\beta_2$, as long as the condition $E_F\gg T$ holds. As such, we neglect $\beta_{1},\beta_2$ at low temperature, leading to a simplified effective free energy
\begin{equation}
	\begin{split}
		\mathcal{F}_{\text{eff}} & =\alpha\sum_i|\Delta_{\bm{Q}_i}|^2+\alpha_1\sum_i|\hat{\bm{Q}}_{i}^\parallel\cdot\boldsymbol{\partial}\Delta_{\bm{Q}_i}|^2 \\&\hspace{0.2in}+\alpha_2\sum_i|\hat{\bm{Q}}_{i}^\perp\cdot\boldsymbol{\partial}\Delta_{\bm{Q}_i}|^2\\&\hspace{0.2in}+\beta\left(\Delta_{\bm{P}}\Delta_{-\bm{P}}\Delta^*_{\bm{Q}}\Delta^*_{-\bm{Q}}+\text{h.c.}\right)+\mathcal{O}(\Delta^6).
	\end{split}\label{feff}
\end{equation}

\subsection{Non-Linear Sigma Model}

We now incorporate phase fluctuations of the PDW bosons through a long-wavelength non-linear sigma model (nLSM) action. This formalism is a first step in providing an explicit determination of the KT transition temperatures into the induced orders, mediated by the proliferation of topological defects, and treats all instabilities in an unbiased manner. The nLSM can be constructed from the GL free energy density Eq.~(\ref{feff}) by assuming the longitudinal fluctuations of the PDW order parameters are frozen, while their phases remain fluctuating as Goldstone modes (additionally assuming the PDW orders are in the type-II limit). This leads to the PDW order parameters
\begin{align}
    \Delta_{\pm \bm{Q}}=|\Delta|\exp\left(i\theta_{\bm{\pm Q}}\right),&&\Delta_{\pm \bm{P}}=|\Delta|\exp\left(i\theta_{\bm{\pm P}}\right),\label{OPsphase}
\end{align}
where the PDW phases are defined through
\begin{align}
    \theta_{\bm{\pm Q}}=\theta\pm\phi,&&\theta_{\bm{\pm P}}=\overline\theta\pm\overline\phi.\label{PDWphases}
\end{align} 
Under $C_4$ symmetries, the phase variables transform as
\begin{align}
    (\theta,\overline{\theta})\xrightarrow{C_4}(\overline\theta,\theta),&& (\phi,\overline{\phi})\xrightarrow{C_4}(\overline\phi,-\phi).
    \label{eq:c4}
\end{align}
The fact that $\phi,\overline{\phi}$ transform the same way as gradients signifies their association with broken translation symmetry. The remaining phases $\theta,\overline{\theta}$ are associated with the breaking of $U(1)$. It is straightforward to see that the phases $2\theta,2\overline{\theta}$ directly correspond to the phases of the composite $4e$-SC order parameters $\Delta_{\bm Q}\Delta_{-\bm Q}$ and $\Delta_{\bm P}\Delta_{-\bm P}$, while $2\phi,2\overline{\phi}$ corresponds to the phases (slide modes) of CDW order parameters $\Psi_{2\bm Q}$ and $\Psi_{2\bm P}$.

By substituting the phase decomposition Eq.~(\ref{OPsphase}) into the effective free energy Eq.~(\ref{feff}), we see that the $\beta$ interaction bears a phase dependence
\begin{equation}
    \mathcal{F}_{\text{eff}}\supset 2\beta |\Delta|^4\cos(2\theta-2\overline{\theta}).
\end{equation}
This interaction represents a Josephson coupling between the phases $2\theta,2\overline{\theta}$ of the $4e$-SC order. Interestingly, even though the $\beta$ interaction can be decomposed in a mean-field theory in both CDW and $4e$-SC channels under a Hubbard-Stratonovich transformation, we see clearly that it only couples to the phase fluctuations in the $4e$-SC channel. The amplitude $|\Delta|$, determined through a minimization of the free energy, depends on the sign of this term. As such, the associated nLSM must assume a fixed difference between the phases of the $4e$-SC order. We note from the dominance of the $\beta$ coefficient at low temperatures that the free energy is biased towards the locking\footnote{We emphasize that despite the $\pm$ sign  on the right hand side of Eq.~\eqref{phaselock}, the sign choice does not represent an Ising-like $\mathbb{Z}_2$ symmetry breaking --- both correspond to the same $d$-wave representation of the $C_4$ symmetry. In fact this $\mathbb{Z}_2$ symmetry can be absorbed into the translation symmetry along the $\pm \bm P$ direction by half the PDW wavelength.} 
\begin{equation}
    2\theta-2\overline{\theta}=\pm\pi\label{phaselock}
\end{equation}
between $4e$-SC phases, which is nothing but a $d$-wave symmetry of the $4e$-SC order, since $2\theta$ and $2\bar\theta$ are related by $C_4$. Under this condition, we minimize $\mathcal{F}_{\rm eff}$, and a substitution of the phase decomposition Eq.~(\ref{OPsphase}) into the free energy results in the nLSM at long wavelengths
\begin{equation}
    \begin{split}
        S&=\beta\int_{\bm{x}}\Big{[}\frac{\kappa_1+\kappa_2}{2}\left|\nabla\theta\right|^2+\frac{1}{2}\left(\kappa_1(\partial_x\phi)^2+\kappa_2(\partial_y\phi)^2\right)\\&\hspace{1in}+\frac{1}{2}\left(\kappa_2(\partial_x\overline{\phi})^2+\kappa_1(\partial_y\overline{\phi})^2\right)\Big{]},
    \end{split}\label{nlsmfinal}
\end{equation}
where we have defined $\int_{\bm{x}}\equiv \int d^2\bm{x}$, $\beta$ the inverse temperature, and we have neglected terms independent of PDW phases. The parameters $\kappa_{1,2}$ descend from the $\alpha_{1,2}$ term in the GL theory. For systems with approximate SO(2) symmetry we have
\begin{equation}
    \kappa_1\gg \kappa_2,
\end{equation}
and we will treat $\kappa_1/\kappa_2$ as a free parameter.
As a consequence of the locking between $4e$-SC phases due to the $\beta$ interaction, the stiffness of $\theta$ receives contributions from both $\Delta_{\pm \bm Q}$ and $\Delta_{\pm \bm P}$, and is isotropic. We will focus on the analysis of the nLSM Eq.~\eqref{nlsmfinal} next.  We note that the symmetry of the system allows for other terms in the action, which we shall discuss in Sec.~\ref{sec:kappa45}.

\subsection{Topological Defects}\label{sec:def}

From the nLSM at long wavelengths, it appears as though the $\theta$- and $\phi$-sectors are independent, but this is not the case. As is well-known, the phase diagram of the nLSM is controlled by both long wavelength physics and, crucially, the topological defects persisting all the way down to lattice scale. In this section, we identify the pertinent topological defects from the physics of the PDW order.

To characterize the topological defects, we introduce the set of fractional/integral topological charges
\begin{equation}
    \begin{split}
        q_\theta&=\frac{n_{\bm{Q}}+n_{-\bm{Q}}}{2}=\frac{n_{\bm{P}}+n_{-\bm{P}}}{2},\\
        q_\phi&=\frac{n_{\bm{Q}}-n_{-\bm{Q}}}{2},\\q_{\overline\phi}&=\frac{n_{\bm{P}}-n_{-\bm{P}}}{2},
    \end{split}\label{charges}
\end{equation}
where $n_{\bm{Q}_i}\in\mathbb{Z}$ is the winding number of the order parameter phase $\theta_{\bm{Q}_i}$ around the defect. It follows that around a topological defect denoted by $q=(q_\theta,q_\phi,q_{\overline\phi})$, the fields have the following holonomy:
\begin{align}
    \begin{split}
        &\theta\to\theta+2\pi q_\theta, \\
    &\phi\to\phi+2\pi q_\phi,\\
    &\overline\phi\to\overline\phi+2\pi q_{\overline\phi}.
    \end{split}\label{windings}
\end{align}
From Eq.~\eqref{charges}, $q_{\theta,\phi,\overline\phi}$ are half integers, and for our purposes,  relevant windings of $\theta$ consist of half $(\pi)$ and full $(2\pi)$ SC vortices, and windings of $\phi,\overline{\phi}$ consist of single $(\pi)$ and double $(2\pi)$ dislocations in the CDW order. Due to the fluctuations of the phase variables in Eq.~\eqref{nlsmfinal}, these defects are subject to long-range Coulomb self-interactions. 

A $2\pi$ winding of a \textit{single} order parameter $\Delta_{\bm{Q}_i}$ results in simultaneous windings of $\pi$ in both $\theta$ and $\phi$-phases, which on the physical level binds half SC vortices and single CDW dislocations together~\cite{annurev:/content/journals/10.1146/annurev-conmatphys-031119-050711}. These half vortex-single dislocations (HVSDs) carry a SC flux quantum of $\Phi_0/2=hc/4e$, and restore both $U(1)$ and translation invariance when proliferated, completely disordering the system. In terms of their topological charges, these HVSDs are denoted by
\begin{equation}   
q=
    \begin{cases}
        \left(\pm \frac{1}{2},\pm\frac{1}{2},0\right),&2\pi\;\text{winding}\;\theta_{\bm{Q}}\\
        \left(\pm\frac{1}{2},\mp \frac{1}{2},0\right),&2\pi\;\text{winding}\;\theta_{-\bm{Q}}\\
        \left(\pm \frac{1}{2},0,\pm\frac{1}{2}\right),&2\pi\;\text{winding}\;\theta_{\bm{P}}\\
        \left(\pm \frac{1}{2},0,\mp \frac{1}{2}\right),&2\pi\;\text{winding}\;\theta_{-\bm{P}}.
    \end{cases}\label{hvsds}
\end{equation}

To next lowest order in topological charge are full vortices, carrying a flux quantum $\Phi_0$, or double CDW dislocations. These defects can be viewed through combinations of HVSDs,
\begin{equation}
    q=\begin{cases}
        \left(\pm 1,0,0\right),&2\pi\;\text{winding}\;\theta\\
        \left(0,\pm 1,0\right),&2\pi\;\text{winding}\;\phi\\
        \left(0,0,\pm1\right),&2\pi\;\text{winding}\;\overline\phi
    \end{cases}\label{fvdd}
\end{equation}
but it is important to note that for particular values of $\kappa_{1,2}$, these defects can be more relevant than HVSDs, in which case they will proliferate independently. Such processes give rise to more novel orders~\cite{Berg2009,radzPRA}: the proliferation of double dislocations restores translation invariance, resulting in a $d$-wave $4e$-SC order, whereas the proliferation of full vortices restores $U(1)$, resulting in a CDW order. It is not necessary to explicitly incorporate the effects of defects with higher-order topological charge, due to a stronger irrelevance under RG.

\subsection{Sine-Gordon Model}\label{sec:SGmodel}

The long-wavelength nLSM, being an expansion in gradients of PDW phases, is ill-equipped at capturing the presence of the topological defects. The interplay between defects characterizes various pathways in which the various symmetries broken in the PDW state are restored, and as such, is crucial in determining the transition temperatures and boundaries between induced orders. For this purpose, it is advantageous to map the nLSM to a dual sine-Gordon model, in which the topological defects are described as local fields.

The mapping of an XY model to its associated sine-Gordon model is well established \cite{Kogut1979,Jose1977,Berg2009,radzPRA}. Compared with the original XY model, our problem involves multiple fields and topological defects, and the stiffness for $\phi$ and $\overline{\phi}$ is anisotropic. Another difference is that, unlike the XY model, the angular variables $\theta,\phi,\overline{\phi}$ do not live on the lattice. Instead, our goal is to map one continuum field theory to another, and we do this by placing the fields on the lattice at an intermediate step. Via this mapping we obtain the local dual operators that correspond to the topological defects in the nLSM, and then we restrict the defect operators to those we previously identified in Sec.~\ref{sec:def}.
The derivation follows closely the well-established procedure for the XY model. For completeness we present it in details, and will show in Sec.~\ref{sec:kappa45} that a similar approach can also be used to treat additional symmetry-allowed terms.

We begin by representing the nLSM Eq.~(\ref{nlsmfinal}) as the continuum limit of an associated 2d XY model, which in our case is described by the partition function
\begin{widetext}

\begin{equation}
    \begin{split}
        Z&=\int\mathcal{D}\theta\mathcal{D}\phi\mathcal{D}\overline{\phi}\prod_{\bm{x}}\exp\Big{(}\frac{\beta(\kappa_1+\kappa_2)}{4}\left(\cos(2\Delta_x\theta)+\cos(2\Delta_y\theta)\right)+\frac{\beta\kappa_1}{4}\left(\cos(2\Delta_x\phi)+\cos(2\Delta_y\overline{\phi})\right)\\&\hspace{3in}+\frac{\beta\kappa_2}{4}\left(\cos(2\Delta_y\phi)+\cos(2\Delta_x\overline{\phi})\right)\Big{)},
    \end{split}
\end{equation}
with $\int\mathcal{D}\theta\equiv \int_0^{2\pi}\prod_{\bm{x}}(d\theta/2\pi)$, and $\Delta_\mu$ as the finite difference operator in direction $\mu$. Note here that we have chosen a periodicity $\pi$ for the angular fields, as we allow for half vortices. 

At sufficiently low temperature, the bond variables of the XY model can be approximated by a sum over Gaussians with the same periodicity,
\begin{equation}
    \begin{split}
        \exp\left(\frac{\beta \kappa_\mu}{4}\cos(2\Delta_\mu\theta(r))\right)&\sim\sum_{w_\mu(r)\in\mathbb{Z}}\exp\left(-\frac{\beta\kappa_\mu}{2}(\Delta_\mu\theta(r)-\pi w_\mu(r))^2\right)\\&\sim\sum_{n_\mu(r)\in\mathbb{Z}} \exp\left(2in_\mu(r)\Delta_\mu\theta(r)-\frac{2n_\mu(r)^2}{\beta\kappa_\mu}\right),
    \end{split}\label{Fourier}
\end{equation}
which is known as the Villain approximation \cite{Villain1975,Jose1977,Kogut1979,Granato1986,Yosefin1985}. Here, $n_\mu(r)$ is an integer valued field on the lattice, and the second step follows from an application of the Poisson summation formula
\begin{equation}
    \sum_{w\in\mathbb{Z}} g(w)=\sum_{n\in\mathbb{Z}}\int _{-\infty}^\infty dx\;e^{-2\pi inx}g(x).\label{psf}
\end{equation}
The convenience of the Villain approximation becomes apparent in light of the partition function 
\begin{equation}
    \begin{split}
        Z&=\int\mathcal{D}\theta\mathcal{D}\phi\mathcal{D}\overline{\phi}\prod_{\bm{x}}\sum_{m_\mu,n_\mu,\overline{n}_\mu \in\mathbb{Z}}\exp\Big{(}2 i(m_\mu\Delta_\mu\theta+n_\mu\Delta_\mu\phi+\overline{n}_\mu\Delta_\mu\overline{\phi})-\frac{2}{\beta(\kappa_1+\kappa_2)}(m_x^2+m_y^2)\\&\hspace{3in}-\frac{2}{\beta\kappa_1}\left(n_x^2+\overline{n}_y^2\right)-\frac{2}{\beta\kappa_2}\left(n_y^2+\overline{n}_x^2\right)\Big{)},
    \end{split}\label{PSFZ}
\end{equation}
as the PDW phases can be readily integrated out. These phases can be viewed as Lagrange multipliers ensuring the constraints $\Delta_\mu m_\mu=0$, which motivates the decomposition via a dual variable $m\in\mathbb{Z}$: $m_\mu=\epsilon_{\mu\nu}\Delta_\nu m$, where $\epsilon_{\mu\nu}$ is the Levi-Civita symbol.  The $m_\mu$ live on bonds, whereas the dual variable $m$ lives on sites of the dual lattice. The same holds for both $n_\mu,\;\overline{n}_\mu$, and we find
\begin{equation}
    \begin{split}
        Z&=\prod_{\bm{x}}\sum_{m,n,\overline{n} \in\mathbb{Z}}\exp\Big{(}-\frac{2}{\beta(\kappa_1+\kappa_2)}\left((\Delta_x m)^2+(\Delta_y m)^2\right)-\frac{2}{\beta\kappa_1}\left((\Delta_yn)^2+(\Delta_x\overline{n})^2\right)-\frac{2}{\beta\kappa_2}\left((\Delta_xn)^2+(\Delta_y\overline{n})^2\right)\Big{)}\\&=\int \mathcal{D}\vartheta \mathcal{D}\varphi \mathcal{D}\overline{\varphi}\prod_{\bm{x}}\sum_{p_\theta, p_\phi, p_{\overline\phi} \in\mathbb{Z}}\exp\Big{(}-2\pi i(p_\theta\vartheta+p_{\phi}\varphi+p_{\overline\phi}\overline{\varphi})-\frac{2}{\beta(\kappa_1+\kappa_2)}\left((\Delta_x\vartheta)^2+(\Delta_y\vartheta)^2\right)\\&\hspace{2.5in}-\frac{2}{\beta\kappa_1}\left((\Delta_y\varphi)^2+(\Delta_x\overline{\varphi})^2\right)-\frac{2}{\beta\kappa_2}\left((\Delta_x\varphi)^2+(\Delta_y\overline{\varphi})^2\right)\Big{)}.
    \end{split}\label{bing}
\end{equation}
\end{widetext}
The second step introduces the dual fields $\vartheta,\varphi,\overline{\varphi}$ through use of the Poisson summation formula, Eq.~(\ref{psf}), and the integration measure is defined through $\int\mathcal{D}\vartheta\equiv \int_{-\infty}^\infty \prod_{\bm{x}}d\vartheta$. 

The integer fields $p_{\theta,\phi,\overline\phi}(r)$ characterize the position and charge of the topological defects of the original nLSM: e.g.,  $p_\theta(r_0)=1$ corresponds to a $\pi$ winding in the original variable $\theta$ around $r_0$. In other words, we have $p_{\theta,\phi,\overline\phi} = 2 q_{\theta,\phi,\overline\phi}$. 
Indeed, the dual fields $\vartheta$, $\varphi$, and $\overline{\varphi}$ mediate  effective Coulomb interactions of the topological defects. Under RG, this effective interaction produces quadratic terms $\sim p^2$ in the action, corresponding to a positive chemical potential to all types of defects, with fugacity of defects with higher charges exponentially suppressed. In addition, as we showed in Sec.~\ref{sec:def}, high-energy physics further imposes the binding of half vortices and dislocations. As a result, the summation over $p_{\theta,\phi,\overline\phi} = 2 q_{\theta,\phi,\overline\phi}$ can be truncated to over those in Eqs.~(\ref{hvsds}, \ref{fvdd}). Assuming  small fugacity, the thermal weight from such a summation can be re-exponentiated, leading to the sine-Gordon action:
\begin{equation}
    \begin{split}
        S_{\text{sG}}&=\int_{x}\Big{[}\frac{T}{2(\kappa_1+\kappa_2)}\left|\nabla\vartheta\right|^2+\frac{T}{2\sqrt{\kappa_1\kappa_2}}\left(|\bm{D}\varphi|^2+|\bm{D}'\overline{\varphi}|^2\right)\\&\hspace{0.5in}-g_{\text{fv}}\cos(2\pi\vartheta)-g_{\text{dd}}(\cos(2\pi\varphi)+\cos(2\pi\overline{\varphi}))\\&\hspace{0.5in}-2g_{\text{hvsd}}\cos(\pi\vartheta)(\cos(\pi\varphi)+\cos(\pi\overline{\varphi}))\Big{]},
    \end{split}\label{sinegordon}
\end{equation}
where for convenience we have rescaled all dual fields down by a factor of 2, and absorbed the anisotropy in the $\varphi,\overline{\varphi}$-sector by defining the derivative
\begin{align}
    \bm{D}=\left(\frac{\kappa_1}{\kappa_2}\right)^{1/4}\hat{x}\partial_x+\left(\frac{\kappa_2}{\kappa_1}\right)^{1/4}\hat{y}\partial_y,\label{covderiv}
\end{align}
with $\bm{D}'$ being related to $\bm{D}$ through the replacement $\kappa_1\leftrightarrow\kappa_2$. Here, $g_{\text{fv}}$, $g_{\text{dd}}$, and $g_{\text{hvsd}}$ are fugacities for full SC vortices, double dislocations, and HVSD's respectively, consistent with the symmetry of the problem. A similar sine-Gordon model has been obtained in Refs.~\cite{Berg2009,Fradkin2015} for a unidirectional PDW state directly on phenomenological basis. 


\section{Vestigial and primary orders}
\label{sec:analysis}

In this section we analyze the dual sine-Gordon model developed in Sec.~\ref{sec:SGmodel}. By computing the scaling dimensions of the local vertex operators representing the topological defects, we explicitly determine the transition temperatures into the ordered states and analyze the resulting phase diagram. We then provide an analysis of various terms omitted by our effective model.

\subsection{Phase Diagram}
By the standard reasoning~\cite{Berg2009,Fradkin2013,Granato1986,Yosefin1985,Jian2021Charge4e,Agterberg2011Charge6e}, phase transitions are driven by the proliferation of the topological defects, which is captured by the cosine terms becoming relevant under RG. 
There are three scaling dimensions for this model, each of which can be determined from correlation functions of vertex operators. From Eq.~\eqref{sinegordon} we have
\begin{equation}
        \begin{split}
           \braket{\vartheta_0\vartheta_{\bm{x}}}&=\frac{\kappa_1+\kappa_2}{2\pi T}\ln|\bm{x}|,\\\braket{\varphi_0\varphi_{\bm{x}}}&=\frac{\sqrt{\kappa_1\kappa_2}}{2\pi T}\ln|\bm{x}'|,
        \end{split} \label{corrs} 
\end{equation}
where $\bm{x}'=\sqrt{\kappa_2/T}\,x\hat{x}+\sqrt{\kappa_1/T}\,y\hat{y}$ incorporates the anisotropy. General correlation functions of vertex operators take the form
\begin{equation}
    \begin{split}
        \braket{e^{n\pi i\vartheta_0}e^{-n\pi i\vartheta_{\bm{x}}}}&=|\bm{x}|^{-n^2\pi(\kappa_1+\kappa_2)/2T}\equiv|\bm{x}|^{-2\Delta_{\vartheta,n}},\\
        \braket{e^{n\pi i\varphi_0}e^{-n\pi i\varphi_{\bm{x}}}}&=|\bm{x}'|^{-n^2\pi\sqrt{\kappa_1\kappa_2}/2T}\equiv|\bm{x}'|^{-2\Delta_{\varphi,n}},
    \end{split}\label{scalingdim}
\end{equation}
where we noted that $\vartheta,\varphi$ are Gaussian free fields, and chose the short distance cutoff to be unity.
For the cosine operators present in our sine-Gordon model Eq.~(\ref{sinegordon}), the scaling dimensions are
\begin{equation}
    \begin{split}
        \Delta_{\text{fv}}&=\Delta_{\vartheta,2}=\frac{\pi}{T}(\kappa_1+\kappa_2),\\\Delta_{\text{dd}}&=\Delta_{\varphi,2}=\frac{\pi}{T}\sqrt{\kappa_1\kappa_2},\\ \Delta_{\text{hvsd}}&=\Delta_{\vartheta,1}+\Delta_{\varphi,1}=\frac{\pi}{4T}(\kappa_1+\kappa_2+\sqrt{\kappa_1\kappa_2}).
    \end{split}
\end{equation} Denoting the scaling dimension of each cosine operator as $\Delta_i$, the beta function for fugacity $g_i$ is $dg_i/dl=(2-\Delta_i)g_i$, to lowest order in fugacity. On their own, these operators become relevant for 
\begin{align}
    \frac{\kappa_1+\kappa_2}{T}&<\frac{2}{\pi},\;\;\;\;\;\text{(for full vortices)}\label{fvss}\\
        \frac{\sqrt{\kappa_1\kappa_2}}{T}&<\frac{2}{\pi},\;\;\;\;\;\text{(for double disc.)}\label{doubledisc}\\
        \frac{\kappa_1+\kappa_2}{T}+\frac{\sqrt{\kappa_1\kappa_2}}{T}&<\frac{8}{\pi}.\;\;\;\;\;\text{(for HVSD)}\label{singledisc}
\end{align}
Moreover, since a full vortex and a double dislocations can be viewed as composites of HVSD's, they proliferate whenever HVSD's do. Using the simple algebraic identity $(\kappa_1+\kappa_2)/2>\sqrt{\kappa_1\kappa_2}$, we see that only Eqs.~(\ref{doubledisc}, \ref{singledisc}) correspond to actual phase transitions, while the proliferation of full vortices are always driven by that of HVSD's. When HVSD's proliferate, both translation symmetry and $U(1)$ symmetry are restored, and the system enters the normal state. When double dislocations proliferate prior to HVSD's do, translation symmetry is restored, and the system enters a $d$-wave $4e$-SC. In this state, the HVSDs become undressed half vortices, which appear as the topological defects of the charge-$4e$ SC. By contrast, there does not exist a vestigial CDW phase in which translation symmetry is broken and $U(1)$ is intact. 

\begin{figure}
    \includegraphics[width=\linewidth]{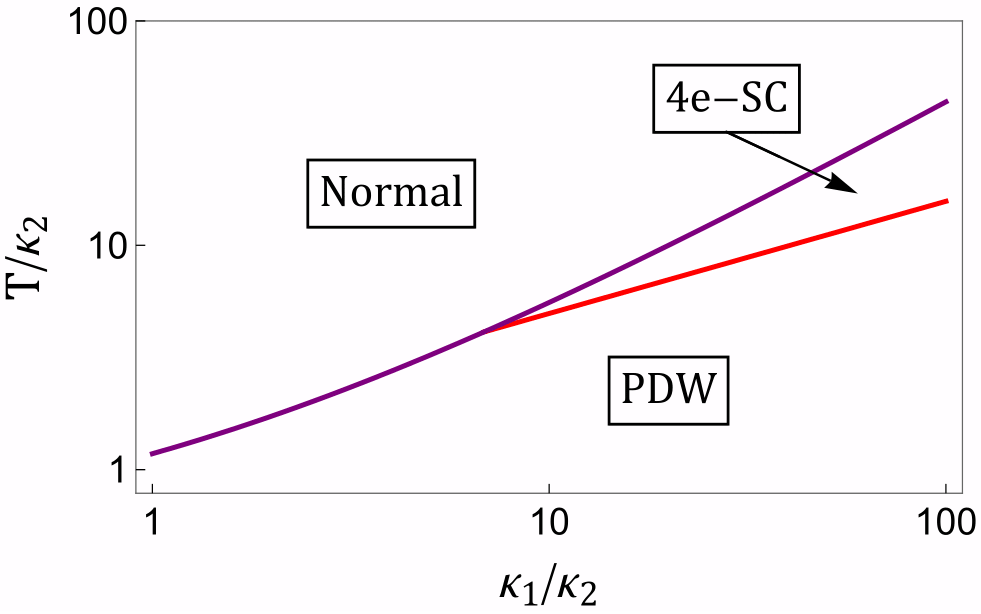}
    \caption{\footnotesize Log-log plot of the phase transitions Eqs. (\ref{doubledisc}-\ref{singledisc}). The red line signifies the proliferation of double CDW dislocations, whereas the purple line signifies the proliferation of HVSDs.}
    \label{fig:phasetrans}
\end{figure}

We depict the phase diagram in Fig.~\ref{fig:phasetrans}. At sufficiently low temperatures, the system is in a PDW state where no topological defects proliferate. For sufficient anisotropy $\kappa_1/\kappa_2\geq (7+3\sqrt{5})/2\approx 6.85$, the system accesses an intermediate phase of $d$-wave charge-$4e$ SC order, in which case double dislocations have proliferated. As the system is pushed to higher temperatures, there is a transition to the normal state through the proliferation of HVSDs, with a direct PDW-disorder transition accessible for smaller values of  $\kappa_1/\kappa_2$. As we mentioned above, for systems with an exact or emergent $SO(2)$ rotation symmetry, $\kappa_1\gg\kappa_2$, and $d$-wave charge-$4e$ SC order exists as an intermediate phase.

\subsection{Additional Interactions}\label{sec:kappa45}

In this section, we discuss additional terms allowed by symmetry beyond those included in the nLSM Eq.~(\ref{nlsmfinal}). The stiffnesses in the nLSM, in principle, incorporate effects from all terms in the GL free energy that are quadratic in gradients of the PDW order parameters. Capturing these effects will result in additional gradient terms that are unrelated to $\kappa_{1,2}$:\begin{equation}
    \begin{split}
        &\delta\kappa|\nabla\theta|^2+\delta\kappa'((\partial_x\phi)^2+(\partial_y\overline{\phi})^2)\\&\hspace{0.25in}+\delta\kappa''((\partial_y\phi)^2+(\partial_x\overline{\phi})^2)
    \end{split}
\end{equation}
From the dominance of the $\beta$ interaction in the GL free energy, it is reasonable to assume coefficients of these interactions to be small compared with $\kappa_{1,2}$. While these terms may alter the exact location of the phase boundaries, we expect the global structure of the phase diagram to be unaffected.

The symmetries of the system, in particular $C_4$ in Eq.~\eqref{eq:c4}, allow for additional cross-gradient terms
\begin{equation}
    \begin{split}
\kappa_4\partial_x\phi\partial_y\overline{\phi}+\kappa_5\partial_y\phi\partial_x\overline{\phi}.
    \end{split}
\end{equation}
While expected to be small, the effect of $\kappa_{4,5}$ terms on the phase diagram is not straightforward. Their treatment in the nLSM is not standard, and we provide a perturbative analysis below. As we will show, regardless of the sign of $\kappa_{4,5}$, they further suppress the tendency towards a vestigial CDW order, making the intermediate $d$-wave $4e$-SC order more robust. 
\begin{widetext}

Focusing on the $\phi,\overline\phi$-sector, we have
\begin{equation}
    \begin{split}
        S_\phi&=\beta\int_x\Big{[}\frac{1}{2}\left(\kappa_1(\partial_x\phi)^2+\kappa_2(\partial_y\phi)^2\right)+\frac{1}{2}\left(\kappa_2(\partial_x\overline{\phi})^2+\kappa_1(\partial_y\overline{\phi})^2\right)+\kappa_4\partial_x\phi\partial_y\overline{\phi}+\kappa_5\partial_y\phi\partial_x\overline{\phi}\Big{]}.
    \end{split}\label{nlsmphi}
\end{equation}
With insight from Eq.~(\ref{Fourier}) in Sec. \ref{sec:SGmodel}, the Villain approximation corresponds to the replacements

\begin{equation}
    \begin{split}
        \exp\left(-\frac{\beta\kappa}{2}(\partial_\mu\phi)^2\right)&\to\sum_{m\in\mathbb{Z}}\exp\left(-\frac{\beta\kappa}{2}(\Delta_\mu\phi-\pi m)^2\right),\\
        \exp\left(-\beta\kappa \partial_x\phi \partial_y \phi\right)&\to\sum_{m_x,m_y\in\mathbb{Z}}\exp\left(-\beta\kappa(\Delta_x\phi-\pi m_x)(\Delta_y\phi-\pi m_y)\right),
    \label{Villain}
    \end{split}
\end{equation}
which incorporate topological defects. The partition function $Z_\phi=\int \mathcal{D}\phi\mathcal{D}\overline\phi\,\exp(-S_\phi)$ can be expressed as
\begin{equation}
    \begin{split}
        Z_\phi&=\int\mathcal{D}\phi\mathcal{D}\overline{\phi}\prod_{\bm{x}}\sum_{m_\mu,\overline{m}_\mu\in\mathbb{Z}}\exp\Big{(}-\frac{\beta}{2}\Big{[}\kappa_1\left((\Delta_x\phi-\pi m_x)^2+(\Delta_y\overline{\phi}-\pi \overline{m}_y)^2\right)+\kappa_2\left((\Delta_y\phi-\pi m_y)^2+(\Delta_x\overline{\phi}-\pi \overline{m}_x)^2\right)\\&\hspace{1.5in}+2\kappa_4(\Delta_x\phi-\pi m_x)(\Delta_y\overline{\phi}-\pi\overline{m}_y )+2\kappa_5(\Delta_y\phi-\pi m_y)(\Delta_x\overline{\phi}-\pi \overline{m}_x)\Big{]}\Big{)},
    \end{split}\label{nlsmphi2}
\end{equation}
\end{widetext}
where we expanded about the global minimum energy configuration set by the dominate $\kappa_{1,2}$ stiffnesses. Conveniently, $Z_{\phi}$ can be expressed in a compact form
\begin{equation}
    \begin{split}
        Z_\phi&=\int\mathcal{D}\phi\mathcal{D}\overline{\phi}\prod_{\bm{x}}\sum_{m_\mu,\overline{m}_\mu\in\mathbb{Z}}\exp\Big{(}-\frac{\beta}{2}\boldsymbol{\Phi}_m^T\boldsymbol{K}\boldsymbol{\Phi}_m\Big{)},
    \end{split}\label{nlsmphi3}
\end{equation}
where the stiffness matrix $\boldsymbol{K}$ and vector of phase variables $\boldsymbol{\Phi}$ are defined through
\begin{align}
    \boldsymbol{K}=\begin{pmatrix}
            \kappa_1&0&0&\kappa_4\\
            0&\kappa_2&\kappa_5&0\\
            0&\kappa_5&\kappa_2&0\\
            \kappa_4&0&0&\kappa_1
        \end{pmatrix},&& \boldsymbol{\Phi}_m=\begin{pmatrix}\Delta_x\phi-\pi m_x\\\Delta_y\phi-\pi m_y\\\Delta_x\overline{\phi}-\pi \overline{m}_x\\\Delta_y\overline{\phi}-\pi \overline{m}_y\end{pmatrix}.
\end{align}

Following the steps shown in Sec. \ref{sec:SGmodel}, with details provided in Appendix \ref{app:k4k5}, this representation for $Z_\phi$ is equivalent to
\begin{equation}
    \begin{split}
        Z_{\varphi}
        &=\int \mathcal{D}\varphi \mathcal{D}\overline{\varphi}\prod_{\bm{x}}\sum_{p,\overline{p}\in\mathbb{Z}}\exp\Big{(}-\pi i(p\varphi+\overline{p}\overline{\varphi})\\&\hspace{1.7in}-\frac{1}{2\beta}\boldsymbol{\Psi}^T\boldsymbol{K}^{-1}\boldsymbol{\Psi}\Big{)},
    \end{split}\label{result}
\end{equation}
in the dual sine-Gordon theory, where the vector of dual fields is defined through
\begin{equation}
    \boldsymbol{\Psi}=\begin{pmatrix}\Delta_y\varphi\\-\Delta_x\varphi\\\Delta_y\overline{\varphi}\\-\Delta_x\overline{\varphi}\end{pmatrix}.
\end{equation}
The leading order correction to the correlation function $\braket{\varphi_0\varphi_{\bm{x}}}$ can be extracted from $Z_\phi$. In the continuum limit, the gradient contribution to the action takes the form 
\begin{equation}
    \begin{split}
        S_{\text{grad}}&=\frac{1}{2\beta}\int \frac{d^2k}{(2\pi)^2}\begin{pmatrix}k_y\varphi\\-k_x\varphi\\k_y\overline{\varphi}\\-k_x\overline{\varphi}\end{pmatrix}^T\boldsymbol{K}^{-1}\begin{pmatrix}k_y\varphi\\-k_x\varphi\\k_y\overline{\varphi}\\-k_x\overline{\varphi}\end{pmatrix}\\
        &\equiv\frac{1}{2}\int \frac{d^2k}{(2\pi)^2}\begin{pmatrix}
            \varphi\\\overline{\varphi}
        \end{pmatrix}^T\boldsymbol{G}^{-1}(\bm{k})\begin{pmatrix}
            \varphi\\\overline{\varphi}
        \end{pmatrix},
    \end{split}
\end{equation}
in momentum space, where we have defined
\begin{equation}
    \begin{split}
        \boldsymbol{G}^{-1}(\bm{k})\equiv\begin{pmatrix}k_y&0\\-k_x&0\\0&k_y\\0&-k_x\end{pmatrix}^T(\beta\boldsymbol{K})^{-1}\begin{pmatrix}k_y&0\\-k_x&0\\0&k_y\\0&-k_x\end{pmatrix}
    \end{split}
\end{equation}
as the inverse Green's function for the dual fields $\varphi,\overline{\varphi}$, or equivalently the inverse of the Fourier transform of the Coulomb interaction between charges of dislocations.

The relevant correlation function then takes the form
\begin{align}
    \braket{\varphi_0\varphi_{\bm{x}}}=\int \frac{d^2k}{(2\pi)^2}\;[\boldsymbol{G}(\bm{k})]_{11}e^{i\bm{k}\cdot\bm{x}}, \label{corrphi}
\end{align}
which can be shown to limit to Eq.~(\ref{corrs}) in the absence of $\kappa_{4,5}$. Since this correlation function is used to determine the scaling dimension of the cosine operators, its renormalization under $\kappa_{4,5}$ can either favor or oppose the proliferation of double dislocations. From the expression of $\boldsymbol{G}^{-1}(\bm{k})$, we find

\begin{equation}
    \begin{split}
        [\boldsymbol{G}(\bm{k})]_{11}&=\frac{\beta\kappa_1\kappa_2}{\kappa_1k_x^2+\kappa_2k_y^2}-\frac{\beta(\kappa_1\kappa_5k_x^2-\kappa_2\kappa_4k_y^2)^2}{(\kappa_2k_x^2+\kappa_1k_y^2)(\kappa_1k_x^2+\kappa_2k_y^2)^2}\\&\hspace{0.5in}+\mathcal{O}(\kappa^3_{4},\kappa^3_5),
    \end{split}\label{G11}
\end{equation}
which when Fourier transformed back to real space Eq. (\ref{corrphi}) allows us to deduce a correction to the scaling dimension Eq. (\ref{scalingdim}) of the cosine operator $\cos(n\pi \varphi)$ to leading order in $\kappa_4,\kappa_5$:
\begin{align}
    \Delta_{\varphi,n}&=\frac{n^2\pi\sqrt{\kappa_1\kappa_2}}{4T}\Big{(}1-\frac{2\kappa_2^2\kappa^2_4+\kappa_1\kappa_2(\kappa_4-\kappa_5)^2+2\kappa_1^2\kappa_5^2}{4\kappa_1\kappa_2(\kappa_1+\kappa_2)^2}\nonumber\\&\hspace{1.5in}+\mathcal{O}(\kappa^3_4,\kappa^3_5)\Big{)}.
\end{align}
It follows that the perturbative inclusion of the $\kappa_4,\kappa_5$ terms lowers the scaling dimension $\Delta_{\varphi,n}$, independent of the sign of $\kappa_{4},\kappa_5$. This further weakens the tendency for the vestigial CDW order. 

\section{Concluding Remarks}
\label{sec:conclusion}

We end this work with a summary of the key aspects of our analysis. Taking a bidirectional PDW model with $C_4$ rotation symmetry, the strong interaction among PDW bosons with orthogonal momenta augments the $4e$-SC stiffness relative to that of the uni-directional model. This augmentation is absent in the CDW sector, and the resulting phase diagram Fig.~\ref{fig:phasetrans} exhibits a $d$-wave charge-$4e$ SC order at intermediate temperatures. 

Our methodology is applicable to a wide range of systems of intertwined orders, with particular relevance to bidirectional systems whose existing studies have been primarily phenomenological. In addition, to the best of our knowledge of the literature of coupled non-linear sigma models, our study of the model Eq.~\eqref{nlsmphi} including mixed gradients and anisotropy is novel. From an experimental perspective, our analysis points to the possibility of $d$-wave $4e$-SC in LBCO at $x=1/8$ doping, with a relative sign change between neighboring Cu-O planes with PDW wave vectors related by $C_4$ rotation \cite{Agterberg2008}.

As an outlook, we note that charge-$6e$ SC has been observed in addition to that of $4e$-SC in the hexagonal lattice structure of $\text{CsV}_3\text{Sb}_5$ \cite{Ge2022Charge4e6e}. Naturally then, it would be interesting to extend our analysis to a PDW state with $C_6$ rotation symmetry. Related to this, recent progress has been made on understanding charge-$6e$ SC from tensor-network approaches \cite{Song2025Phase,Lin2024Theory}, which are a powerful numerical tool in the study of 2D XY models \cite{Vanderstraeten2019_ApproachingKT,SongZhang2022_PhaseCoherence}. It would be interesting to perform a tensor-network study of our associated 2D XY model, where the coupling between the phases of the $4e$-SC order is incorporated explicitly. More recently, an alternative scenario in which translation symmetry is destroyed through fractionalization has been proposed \cite{May-Mann2025}, and it would be insightful to understand how this fractionalization is mediated by topological defects, which we leave for future work.

\begin{acknowledgments}
We would like to thank E. Fradkin, L. Radzihovsky,  Z. Han and Y.-M. Wu for useful discussions. This work was supported by NSF under award number DMR-2045781. 
\end{acknowledgments}

\begin{widetext}
    
\appendix

\section{Details of the Ginzburg-Landau Theory}\label{app:GL}

In this section we will explicitly evaluate the coefficients of quartic interactions
\begin{equation}
    \begin{split}
        &\beta_1\sum_i|\Delta_{\bm{Q}_i}|^4+\beta_2\left(|\Delta_{\bm{Q}}|^2|\Delta_{\bm{P}}|^2+|\Delta_{-\bm{Q}}|^2|\Delta_{\bm{P}}|^2+|\Delta_{\bm{Q}}|^2|\Delta_{-\bm{P}}|^2+|\Delta_{-\bm{Q}}|^2|\Delta_{-\bm{P}}|^2\right)\\&\hspace{0.75in}+\beta\left(\Delta_{\bm{P}}\Delta_{-\bm{P}}\Delta^*_{\bm{Q}}\Delta^*_{-\bm{Q}}+\text{h.c.}\right)
    \end{split}
\end{equation}
in the GL free energy Eq.~(\ref{fmin}). The relevant diagrams are shown in Fig.~\ref{fig:diagrams},
\begin{figure}[h]
    \centering
    \begin{minipage}{0.3\textwidth}
        \centering
        \begin{tikzpicture}[baseline=(o.base)]
    \begin{feynhand}
      \vertex[dot] (a) at (-0.75, -0.75) {};
      \vertex[dot] (b) at (-0.75, 0.75) {};
      \vertex[dot] (c) at (0.75, -0.75) {};
      \vertex[dot] (d) at (0.75, 0.75) {};
      \vertex (e) at (-1.5,-1.5) {};
      \vertex (f) at (-1.5,1.5) {};
      \vertex (g) at (1.5,-1.5) {};
      \vertex (h) at (1.5,1.5) {};
      \vertex (i) at (-1.65,-1.65) {$\Delta^*_{\bm{Q}}$};
      \vertex (j) at (-1.65,1.65) {$\Delta_{\bm{Q}}$};
      \vertex (k) at (1.65,-1.65) {$\Delta_{\bm{Q}}$};
      \vertex (l) at (1.65,1.65) {$\Delta^*_{\bm{Q}}$};
      \vertex (m) at (-1.3,0) {\footnotesize$\bm{k}$};
      \vertex (n) at (1.3,0) {\footnotesize$\bm{k}$};
      \vertex (o) at (0,1.2) {\footnotesize$-\bm{k}-\bm{Q}$};
      \vertex (p) at (0,-1.2) {\footnotesize$-\bm{k}-\bm{Q}$};
      \propag [fermion] (b) to  (a); 
      \propag [fermion] (c) to  (a);
      \propag [fermion] (b) to  (d); 
      \propag [fermion] (c) to  (d);
      \propag [photon] (e) to (a);
      \propag [photon] (f) to (b);
      \propag [photon] (g) to (c);
      \propag [photon] (h) to (d);
    \end{feynhand}
    \end{tikzpicture}
        \caption*{(i)}
    \end{minipage}
    \hfill
    \begin{minipage}{0.3\textwidth}
        \centering
        \begin{tikzpicture}[baseline=(o.base)]
    \begin{feynhand}
      \vertex[dot] (a) at (-0.75, -0.75) {};
      \vertex[dot] (b) at (-0.75, 0.75) {};
      \vertex[dot] (c) at (0.75, -0.75) {};
      \vertex[dot] (d) at (0.75, 0.75) {};
      \vertex (e) at (-1.5,-1.5) {};
      \vertex (f) at (-1.5,1.5) {};
      \vertex (g) at (1.5,-1.5) {};
      \vertex (h) at (1.5,1.5) {};
      \vertex (i) at (-1.65,-1.65) {$\Delta^*_{\bm{P}}$};
      \vertex (j) at (-1.65,1.65) {$\Delta_{\bm{P}}$};
      \vertex (k) at (1.65,-1.65) {$\Delta_{\bm{Q}}$};
      \vertex (l) at (1.65,1.65) {$\Delta^*_{\bm{Q}}$};
      \vertex (m) at (-1.3,0) {\footnotesize$\bm{k}$};
      \vertex (n) at (1.7,0) {\footnotesize$\bm{k}+\bm{P}-\bm{Q}$};
      \vertex (o) at (0,1.2) {\footnotesize$-\bm{k}-\bm{P}$};
      \vertex (p) at (0,-1.2) {\footnotesize$-\bm{k}-\bm{P}$};
      \propag [fermion] (b) to  (a); 
      \propag [fermion] (c) to  (a);
      \propag [fermion] (b) to  (d); 
      \propag [fermion] (c) to  (d);
      \propag [photon] (e) to (a);
      \propag [photon] (f) to (b);
      \propag [photon] (g) to (c);
      \propag [photon] (h) to (d);
    \end{feynhand}
    \end{tikzpicture}
        \caption*{(ii)}
    \end{minipage}
    \hfill
    \begin{minipage}{0.3\textwidth}
        \centering
        \begin{tikzpicture}[baseline=(o.base)]
    \begin{feynhand}
      \vertex[dot] (a) at (-0.75, -0.75) {};
      \vertex[dot] (b) at (-0.75, 0.75) {};
      \vertex[dot] (c) at (0.75, -0.75) {};
      \vertex[dot] (d) at (0.75, 0.75) {};
      \vertex (e) at (-1.5,-1.5) {};
      \vertex (f) at (-1.5,1.5) {};
      \vertex (g) at (1.5,-1.5) {};
      \vertex (h) at (1.5,1.5) {};
      \vertex (i) at (-1.65,-1.65) {$\Delta^*_{\bm{Q}}$};
      \vertex (j) at (-1.65,1.65) {$\Delta_{\bm{P}}$};
      \vertex (k) at (1.65,-1.65) {$\Delta_{-\bm{P}}$};
      \vertex (l) at (1.65,1.65) {$\Delta^*_{-\bm{Q}}$};
      \vertex (m) at (-1.3,0) {\footnotesize$\bm{k}$};
      \vertex (n) at (1.7,0) {\footnotesize$\bm{k}+\bm{P}+\bm{Q}$};
      \vertex (o) at (0,1.2) {\footnotesize$-\bm{k}-\bm{P}$};
      \vertex (p) at (0,-1.2) {\footnotesize$-\bm{k}-\bm{Q}$};
      \propag [fermion] (b) to  (a); 
      \propag [fermion] (c) to  (a);
      \propag [fermion] (b) to  (d); 
      \propag [fermion] (c) to  (d);
      \propag [photon] (e) to (a);
      \propag [photon] (f) to (b);
      \propag [photon] (g) to (c);
      \propag [photon] (h) to (d);
    \end{feynhand}
    \end{tikzpicture}
        \caption*{(iii)}
    \end{minipage}

    \caption{(i), (ii), (iii) are diagrams representing the GL coefficients coefficients $\beta_1$, $\beta_2$, and $\beta$ respectively.}
    \label{fig:diagrams}
\end{figure}
and can be evaluated at low temperature through
\begin{align}
    \beta_1&\sim\int d\omega\int d^2k\;\left(\frac{1}{i\omega-\varepsilon_{\bm{k}}}\right)^2\left(\frac{1}{-i\omega-\varepsilon_{-\bm{k}-\bm{Q}}}\right)^2,\\
    \beta_2&\sim\int d\omega\int d^2k\;\frac{1}{i\omega-\varepsilon_{\bm{k}}}\left(\frac{1}{-i\omega-\varepsilon_{-\bm{k}-\bm{P}}}\right)^2\frac{1}{i\omega-\varepsilon_{\bm{k}+\bm{P}-\bm{Q}}},\\
    \beta&\sim\int d\omega\int d^2k\;\frac{1}{i\omega-\varepsilon_{\bm{k}}}\frac{1}{-i\omega-\varepsilon_{-\bm{k}-\bm{Q}}}\frac{1}{-i\omega-\varepsilon_{-\bm{k}-\bm{P}}}\frac{1}{i\omega-\varepsilon_{\bm{k}+\bm{P}+\bm{Q}}},
\end{align}
where since only the scaling of each coefficient is relevant for our purposes, we will drop various numerical coefficients. For reference, analogous diagrams have been evaluated in a work \cite{Wang2014CDW}.

In the continuum, the ratio $|\bm{Q}|/k_F$ is tunable \cite{PDWorderelecrepul2023}. In the case that $|\bm{Q}|=\sqrt{2}k_F$, the internal fermions lie on the Fermi surface, as shown in Fig.~\ref{fig:FS} of the main text. As we will show, this choice provides the enhancement to the $\beta$ coefficient. The fermionic dispersions can be expanded around these hotspots, confined to a UV energy scale $E_F\gg T $. For $\beta_1$, this results in
\begin{equation}
    \begin{split}
        \beta_1&\sim\frac{1}{v_F^2}\int d\omega\int_{-E_F}^{E_F}\;\frac{dxdy}{(i\omega-x)^2(i\omega-y)^2},
    \end{split}
\end{equation}
where we made a change of variables $x=v_Fk_x$, $y=v_Fk_y$. The integrations are now straightforward,
\begin{equation}
    \begin{split}
        \beta_1&\sim\frac{1}{v_F^2}\int_T d\omega\;\frac{E_F^2}{(E_F^2+\omega^2)^2}\;\;\;\iff\;\;\;\beta_1\sim \frac{1}{v_F^2E_F},
    \end{split}
\end{equation}
where we retained a finite $T$ to regulate the IR singularity, only setting $T\to0 $ at the very end. The $\beta_2$ coefficient follows similarly,
\begin{equation}
    \begin{split}
        \beta_2\sim-\frac{1}{v_F^2}\int d\omega\int_{-E_F}^{E_F}\;\frac{dxdy}{(\omega^2+x^2)(i\omega-y)^2}&\sim\frac{1}{v_F^2}\int _Td\omega\;\frac{E_F}{\omega(E_F^2+\omega^2)}\;\;\;\iff\;\;\;\beta_2\sim \frac{1}{v_F^2E_F}\ln\left(\frac{E_F}{T}\right),
    \end{split}
\end{equation}
and bears a logarithmic singularity as $T\to0$. Last, the $\beta$ coefficient scales as
\begin{equation}
    \begin{split}
        \beta\sim\frac{1}{v_F^2}\int d\omega\int_{-E_F}^{E_F}\;\frac{dxdy}{(\omega^2+x^2)(\omega^2+y^2)}&\sim\frac{1}{v_F^2}\int_T d\omega\;\frac{1}{\omega^2}\;\;\;\iff\;\;\;\beta\sim \frac{1}{v_F^2T},
    \end{split}
\end{equation}
where we extended the limits of momentum integration to infinity since the divergence comes from momenta much smaller than $E_F$. It is clear to see that the $\beta$ interaction is dominate in the regime $E_F\gg T$.

\section{Details of the Analysis with \texorpdfstring{\boldmath$\kappa_{4},\kappa_5$}{\kappa_{4},\kappa_5} Interactions}\label{app:k4k5}

In this section we show how the representation Eq.~(\ref{result})
\begin{equation}
    \begin{split}
        Z_{\varphi}
        &=\int \mathcal{D}\varphi \mathcal{D}\overline{\varphi}\prod_{\bm{x}}\sum_{p,\overline{p}\in\mathbb{Z}}\exp\Big{(}-\pi i(p\varphi+\overline{p}\overline{\varphi})-\frac{1}{2\beta}\boldsymbol{\Psi}^T\boldsymbol{K}^{-1}\boldsymbol{\Psi}\Big{)}\\&=\int \mathcal{D}\varphi \mathcal{D}\overline{\varphi}\prod_{\bm{x}}\sum_{p,\overline{p}\in\mathbb{Z}}\exp\Big{(}-\pi i(p\varphi+\overline{p}\overline{\varphi})-\frac{1}{2\beta}\begin{pmatrix}\Delta_y\varphi\\-\Delta_x\varphi\\\Delta_y\overline{\varphi}\\-\Delta_x\overline{\varphi}'\end{pmatrix}^T\begin{pmatrix}
            \kappa_1&0&0&\kappa_4\\
            0&\kappa_2&\kappa_5&0\\
            0&\kappa_5&\kappa_2&0\\
            \kappa_4&0&0&\kappa_1
        \end{pmatrix}^{-1}\begin{pmatrix}\Delta_y\varphi\\-\Delta_x\varphi\\\Delta_y\overline{\varphi}\\-\Delta_x\overline{\varphi}\end{pmatrix}\Big{)},
    \end{split}\label{resultA}
\end{equation}
for the partition function of the dual $\varphi,\overline{\varphi}$ fields follows from the partition function of the original $\phi,\overline{\phi}$ phases Eq.~(\ref{nlsmphi3})
\begin{equation}
    \begin{split}
        Z_\phi&=\int\mathcal{D}\phi\mathcal{D}\overline{\phi}\prod_{\bm{x}}\sum_{m_\mu,\overline{m}_\mu\in\mathbb{Z}}\exp\Big{(}-\frac{\beta}{2}\boldsymbol{\Phi}^T\boldsymbol{K}\boldsymbol{\Phi}\Big{)}\\&=\int\mathcal{D}\phi\mathcal{D}\overline{\phi}\prod_{\bm{x}}\sum_{m_\mu,\overline{m}_\mu\in\mathbb{Z}}\exp\Big{(}-\frac{\beta}{2}\begin{pmatrix}\Delta_x\phi-\pi m_x\\\Delta_y\phi-\pi m_y\\\Delta_x\overline{\phi}-\pi \overline{m}_x\\\Delta_y\overline{\phi}-\pi \overline{m}_y\end{pmatrix}^T\begin{pmatrix}
            \kappa_1&0&0&\kappa_4\\
            0&\kappa_2&\kappa_5&0\\
            0&\kappa_5&\kappa_2&0\\
            \kappa_4&0&0&\kappa_1
        \end{pmatrix}\begin{pmatrix}\Delta_x\phi-\pi m_x\\\Delta_y\phi-\pi m_y\\\Delta_x\overline{\phi}-\pi \overline{m}_x\\\Delta_y\overline{\phi}-\pi \overline{m}_y\end{pmatrix}\Big{)},
    \end{split}\label{nlsmphi3A}
\end{equation}
shown in Section \ref{sec:kappa45} of the main text. 

First, the application of the Poisson summation formula (PSF)
\begin{equation}
    \sum_{n\in\mathbb{Z}} g(n)=\sum_{m\in\mathbb{Z}}\int dx\;e^{-2\pi imx}g(x), \label{psf2}
\end{equation}
on Eq.~(\ref{nlsmphi3A}) results in the intermediate form
\begin{equation}
    \begin{split}
        Z_\phi&=\int\mathcal{D}\phi\mathcal{D}\overline{\phi}\prod_{\bm{x}}\sum_{n_\mu,\overline{n}_\mu\in\mathbb{Z}}\exp\Big{(}-2 i(n_\mu\Delta_\mu\phi+\overline{n}_\mu\Delta_\mu\overline{\phi})-\frac{2}{\beta}\begin{pmatrix}n_x\\n_y\\\overline{n}_x\\\overline{n}_y\end{pmatrix}^T\begin{pmatrix}
            \kappa_1&0&0&\kappa_4\\
            0&\kappa_2&\kappa_5&0\\
            0&\kappa_5&\kappa_2&0\\
            \kappa_4&0&0&\kappa_1
        \end{pmatrix}^{-1}\begin{pmatrix}n_x\\n_y\\\overline{n}_x\\\overline{n}_y\end{pmatrix}\Big{)},
    \end{split}\label{nlsmphi4}
\end{equation}
which is the analogous representation of Eq.~(\ref{PSFZ}) in the main text when $\kappa_4,\kappa_5$ interactions are present. The proof of this fact is shown in Sec. \ref{proofsec} below. The next step is to integrate out $\phi,\overline{\phi}$, leading to the constraints $\Delta_\mu n_\mu=\Delta_\mu\overline{n}_\mu=0$ which justify the decompositions $n_\mu=(\Delta_yn,-\Delta_xn)$, $\overline{n}_\mu=(\Delta_y\overline{n},-\Delta_x\overline{n})$. What results is
\begin{equation}
    \begin{split}
        Z_\varphi&=\prod_{\bm{x}}\sum_{n,\overline{n}\in\mathbb{Z}}\exp\Big{(}-\frac{2}{\beta}\begin{pmatrix}\Delta_yn\\-\Delta_xn\\\Delta_y\overline{n}\\-\Delta_x\overline{n}\end{pmatrix}^T\begin{pmatrix}
            \kappa_1&0&0&\kappa_4\\
            0&\kappa_2&\kappa_5&0\\
            0&\kappa_5&\kappa_2&0\\
            \kappa_4&0&0&\kappa_1
        \end{pmatrix}^{-1}\begin{pmatrix}\Delta_yn\\-\Delta_xn\\\Delta_y\overline{n}\\-\Delta_x\overline{n}\end{pmatrix}\Big{)}\\
        &=\int \mathcal{D}\varphi \mathcal{D}\overline{\varphi}\prod_{\bm{x}}\sum_{p,\overline{p}\in\mathbb{Z}}\exp\Big{(}-\pi i(p\varphi+\overline{p}\overline{\varphi})-\frac{1}{2\beta}\begin{pmatrix}\Delta_y\varphi\\-\Delta_x\varphi\\\Delta_y\overline{\varphi}\\-\Delta_x\overline{\varphi}\end{pmatrix}^T\begin{pmatrix}
            \kappa_1&0&0&\kappa_4\\
            0&\kappa_2&\kappa_5&0\\
            0&\kappa_5&\kappa_2&0\\
            \kappa_4&0&0&\kappa_1
        \end{pmatrix}^{-1}\begin{pmatrix}\Delta_y\varphi\\-\Delta_x\varphi\\\Delta_y\overline{\varphi}\\-\Delta_x\overline{\varphi}\end{pmatrix}\Big{)},
    \end{split}\label{nlsmphi5}
\end{equation}
where the second step made an additional use of the Poisson summation formula, and was followed by a field redefinition to manifest the mod $\pi$ periodicity of the original PDW phases. The dual representation Eq.~(\ref{nlsmphi5}) is precisely that shown in the main text Eq.~(\ref{result}).

\subsection{Proof of Intermediate Representation Eq.~(\ref{nlsmphi4})}\label{proofsec}

\hspace{0.125in} We first decompose $Z_\phi$ through
\begin{equation}
Z_\phi=\int\mathcal{D}\phi\mathcal{D}\overline{\phi}\prod_{\vb{x}}f_{\kappa_4}[\phi,\overline{\phi}]f_{\kappa_5}[\phi,\overline{\phi}],
\end{equation}
where $f_{\kappa_4}[\phi,\overline\phi]$ takes the form
\begin{equation}
    f_{\kappa_4}[\phi,\overline{\phi}]=\sum_{m_x,\overline{m}_y}\exp\left(-\frac{\beta\kappa_1}{2}(\Delta_x\phi-\pi m_x)^2-\beta\kappa_4(\Delta_x\phi-\pi m_x)(\Delta_y\overline{\phi}-\pi \overline{m}_y)-\frac{\beta\kappa_1}{2}(\Delta_y\overline{\phi}-\pi \overline{m}_y)^2\right),\label{wasB6}
\end{equation}
and $f_{\kappa_5}[\phi,\overline\phi]$ is analogously defined. For brevity, we will investigate only the $\kappa_4$ sector, which involves the variables $m_x$ and $\overline{m}_y$, as the $\kappa_5$ sector follows from the same methods. We first apply the PSF Eq.~(\ref{psf2}) on the $m_x$ variable in Eq.~(\ref{wasB6}),
\begin{equation}
    \begin{split}
        f_{\kappa_4}[\phi,\overline{\phi}]&=\sum_{n_x,\overline{m}_y}\int_x\exp\left(-2\pi in_xx-\frac{\beta\kappa_1}{2}(\Delta_x\phi-\pi x)^2-\beta\kappa_4(\Delta_x\phi-\pi x)(\Delta_y\overline{\phi}-\pi \overline{m}_y)-\frac{\beta\kappa_1}{2}(\Delta_y\overline{\phi}-\pi \overline{m}_y)^2\right)\\&=\sum_{n_x,\overline{m}_y}\exp\left(-2in_x\Delta_x\phi-\frac{\beta\kappa_1}{2}(\Delta_y\overline{\phi}-\pi \overline{m}_y)^2\right)\int_y\exp\left((2\pi in_x-\pi\beta\kappa_4(\Delta_y\overline{\phi}-\pi \overline{m}_y))y-\frac{\beta\kappa_1}{2}(\pi y)^2\right)\\&=\sum_{n_x,\overline{m}_y}\exp\left(-2in_x\Delta_x\phi-\frac{2n_x^2}{\beta\kappa_1}\right)\exp\Big{(}-\frac{2i\kappa_4n_x}{\kappa_1}(\Delta_y\overline{\phi}-\pi\overline{m}_y)-\frac{\beta}{2}\left(\kappa_1-\frac{\kappa_4^2}{\kappa_1}\right)(\Delta_y\overline{\phi}-\pi \overline{m}_y)^2\Big{)},
    \end{split}
\end{equation}
where the second equality follows a shift of the integration variable $\pi y=\Delta_x\phi-\pi x$. We now follow the same procedure for the remaining variable $\overline{m}_y$. Implementing the PSF, we find
\begin{equation}
    \begin{split}
        f_{\kappa_4}[\phi,\overline{\phi}]&=\sum_{n_x,\overline{n}_y}\exp\left(-2in_x\Delta_x\phi-\frac{2n_x^2}{\beta\kappa_1}\right)\int_x\exp\Big{(}-2\pi i\overline{n}_yx-\frac{2i\kappa_4n_x}{\kappa_1}(\Delta_y\overline{\phi}-\pi x)-\frac{\beta}{2}\left(\kappa_1-\frac{\kappa_4^2}{\kappa_1}\right)(\Delta_y\overline{\phi}-\pi x)^2\Big{)}\\&=\sum_{n_x,\overline{n}_y}\exp\left(-2i(n_x\Delta_x\phi+\overline{n}_y\Delta_y\overline{\phi})-\frac{2}{\beta\kappa_1\left(1-\kappa_4^2/\kappa_1^2\right)}\left(n_x^2+\overline{n}_y^2-\frac{2\kappa_4}{\kappa_1}n_x\overline{n}_y\right)\right),
    \end{split}\label{alj}
\end{equation}
where the second equality follows from a shift of variable $\pi y=\Delta_y\overline{\phi}-\pi\overline{m}_y$ and evaluation of the resulting Gaussian.

The oscillatory term generated in Eq.~(\ref{alj}) is seen directly in the desired relation Eq.~(\ref{nlsmphi4}), and by expanding the matrix product in Eq.~(\ref{nlsmphi4}), it is easy to see that
\begin{equation}
    \begin{split}
        -\frac{2}{\beta}\begin{pmatrix}n_x\\n_y\\\overline{n}_x\\\overline{n}_y\end{pmatrix}^T\begin{pmatrix}
            \kappa_1&0&0&\kappa_4\\
            0&\kappa_2&\kappa_5&0\\
            0&\kappa_5&\kappa_2&0\\
            \kappa_4&0&0&\kappa_1
        \end{pmatrix}^{-1}\begin{pmatrix}n_x\\n_y\\\overline{n}_x\\\overline{n}_y\end{pmatrix}&=-\frac{2}{\beta}\begin{pmatrix}n_x\\n_y\\\overline{n}_x\\\overline{n}_y\end{pmatrix}^T\begin{pmatrix}
            \frac{\kappa_1}{\kappa_1^2-\kappa_4^2}&0&0&-\frac{\kappa_4}{\kappa_1^2-\kappa_4^2}\\
            0&\cdots&\cdots&0\\
            0&\cdots&\cdots&0\\
            -\frac{\kappa_4}{\kappa_1^2-\kappa_4^2}&0&0&\frac{\kappa_1}{\kappa_1^2-\kappa_4^2}
        \end{pmatrix}\begin{pmatrix}n_x\\n_y\\\overline{n}_x\\\overline{n}_y\end{pmatrix}\\&=-\frac{2}{\beta\kappa_1(1-\kappa_4^2/\kappa_1^2)}\left(n_x^2+\overline{n}^2_y-\frac{2\kappa_4}{\kappa_1}n_x\overline{n}_y\right)+\cdots,
    \end{split}
\end{equation}
which is the remaining term in Eq.~(\ref{alj}). The ''$\cdots$" are terms involving $\kappa_5$, and the same method can be applied in that sector. 

\end{widetext}

\bibliography{4ephase}

\end{document}